\documentclass[12pt,preprint]{aastex}

\shorttitle{Proto-Neutron and Neutron Stars in a Chiral SU(3) Model}
\shortauthors{Dexheimer et al.}

\begin{document}


\title{Proto-Neutron and Neutron Stars in a Chiral SU(3) Model}


\author{V. Dexheimer and S. Schramm}
\affil{Frankfurt Institute for Advanced Studies\\Johann Wolfgang Goethe University\\
Ruth-Moufang-Str. 1, 60438 Frankfurt am Main, Germany.}
\email{dexheimer@th.physik.uni-frankfurt.de}



\begin{abstract}
A hadronic chiral SU(3) model is applied to neutron and proto-neutron
stars, taking into account trapped neutrinos, finite
temperature and entropy.
The transition to the chirally restored phase is
studied and global properties of the stars like
minimum and maximum masses and radii are calculated for different
cases. In addition, the effects of rotation on
neutron star masses are included and the conservation of baryon number and angular momentum determine the maximum frequencies of rotation during the cooling.
\end{abstract}





\section{Introduction}

After a core-collapse supernova explosion of a star with a mass smaller than about 20 solar masses,
the remaining star, initially called proto-neutron star, is left with a very high temperature of up to $50$ MeV. Due to their short
mean free path especially electron neutrinos stay trapped for roughly 10 to 20 seconds and develop a finite chemical potential.
In this stage the structure of the star can be divided into a core region that will be modelled in this article,
and an envelope with higher entropy. Maximum and minimum star masses as well as rotational constraints on them will be considered.
Finally, the corresponding properties of the cooled neutron star will be determined as well.

As a direct solution of the QCD equations for a high-density
hadronic environment is currently out of reach, the core region of
the star will be described using an effective SU(3) chiral model
including hyperonic degrees of freedom. The effect of strange hadrons on the properties of neutron stars has been discussed in the past by a number of authors including the investigation of kaon condensation (\citet{kaplan,kcond}) and studies of the effect of hyperons in neutron stars (\citet{glen1,glen2};\citet{hyperons,amruta}) and references in \citet{review}).
Here in a general approach the baryons and mesonic
fields are introduced as flavor-SU(3) multiplets. The baryons
obtain their masses through their coupling to the scalar meson
fields via spontaneous symmetry breaking. The parameters of the
theory are fixed to reproduce the hadronic vacuum masses as well
as the properties of ground state nuclear matter. The model is
extended to include effects from the lowest spin-3/2 baryonic
multiplet. The model has been successfully applied to the
description of nuclear matter and finite nuclei (\citet{chiral1,
chiral2}) as well as to the high-temperature and high-density
environment of an ultrarelativistic heavy-ion collision
(\citet{phasediagram}). A study of neutron stars had also been
performed (\citet{neutronSchramm, neutronSchramm2}). In the
following these calculations are extended to proto-neutron star
environments with finite temperature and neutrino chemical
potential, additionally considering the population of higher baryonic
resonances. In a further refitting of the basic model parameters,
a study of the maximally achievable star masses in this
approach is presented, while still maintaining a phenomenologically reasonable
description of ground state nuclear matter.

The article is organized as follows. In section 2 the hadronic degrees of freedom and the model Lagrangian will be briefly presented. In section 3 the
model will be applied to zero-temperature neutron stars with special emphasis on the effect of non-linear meson interactions.
The properties of proto-neutron stars at finite temperature, entropy and lepton number are shown and discussed in section 4. Finally, the effects of
rotation and its consequences on the cooling of the star are analyzed and conclusions are presented.

\section{The Hadronic Model}

Since in a high-density environment like the one present in the
center of neutron stars the baryonic chemical potential is high
enough to create particles beyond the lowest SU(2) multiplet of
the nucleons, which comprise hyperons ($\Lambda$, $\Sigma$, $\Xi$) and
possibly other resonance states ($\Delta$, $\Sigma^*$, $\Xi^*$,
$\Omega$), the hadronic model is based on a flavor-SU(3)
description including hyperons and strange mesons as part of the
basic hadronic SU(3) multiplets, which were also partially taken into
account in previous works (\citet{jurgen, amruta}). The baryons
interact via exchange of scalar ($\sigma$, $\delta$, $\zeta$,
$\chi$) and vector mesons ($\rho$, $\omega$, $\phi$). The
isovector vector ($\rho$) and scalar ($\delta$) mesons are
included in a natural way, as an essential ingredient for
reproducing the phenomenological value for the asymmetry energy of
nuclear matter. A scalar isoscalar (dilaton) field ($\chi$) acts
as an effective gluon condensate. On the leptonic side, electrons
and muons assure charge neutrality and neutrinos are considered in
the case of proto-neutron stars.

The Lagrangian density of the chiral model in mean field
approximation used in our calculations reads:
\begin{eqnarray}
&L_{MFT}=L_{Kin}+L_{Bscal}+L_{Bvec}+L_{scal}+L_{vec}+L_{SB},&
\end{eqnarray}
where besides the kinetic energy term for baryons and leptons, the terms:
\begin{eqnarray}
&L_{Bscal}+L_{Bvec}=-\sum_i \bar{\psi_i}[g_{i\omega}\gamma_0\omega+g_{i\phi}\gamma_0\phi+g_{i\rho}\gamma_0\tau_3\rho+m_i^*]\psi_i,&
\end{eqnarray}
\begin{eqnarray}
&L_{vec}=-\frac{1}{2}(m_\omega^2\omega^2+m_\rho^2\rho^2+m_\phi^2\phi^2)\frac{\chi^2}{\chi_o^2}
+ L_{vec,4},&
\end{eqnarray}
\begin{eqnarray}
&L_{scal}=\frac{1}{2}k_0\chi^2(\sigma^2+\zeta^2+\delta^2)-k_1(\sigma^2+\zeta^2+\delta^2)^2&\nonumber\\&-k_2\left(\frac{\sigma^4}{2}+\frac{\delta^4}{2}
+3\sigma^2\delta^2+\zeta^4\right)
-k_3\chi(\sigma^2-\delta^2)\zeta&\nonumber\\&+k_4\chi^4+\frac{1}{4}\chi^4\ln{\frac{\chi^4}{\chi_0^4}}-
\epsilon\ \ \chi^4\ln{\frac{(\sigma^2-\delta^2)\zeta}{\sigma_0^2\zeta_0}},&
\end{eqnarray}
\begin{eqnarray}
&L_{SB}=\left(\frac{\chi}{\chi_0}\right)^2\Bigg[m_\pi^2 f_\pi\sigma+\left(\sqrt{2}m_k^ 2f_k-\frac{1}{\sqrt{2}}m_\pi^ 2 f_\pi\right)\zeta\Bigg],&
\end{eqnarray}
represent the interactions between baryons and scalar mesons and between baryons and vector
mesons, the self interactions of scalar and vector mesons and an explicitly chiral symmetry breaking term, responsible for
producing the masses of the pseudo-scalar mesons. In Eqs. (2-5) the mesons are treated as classical fields within the mean-field
approximation. Only the $0^{th}$ component of the vector meson survives, i.e. $\omega\equiv\omega_0$, $\rho\equiv\rho_0^0$, $\phi\equiv\phi_0$.
$\sigma$ and $\zeta$ are the scalar fields corresponding to the non-strange and strange quark condensate, respectively.  A detailed
discussion of this Langrangian can be found in \citet{chiral1,chiral2}.
The baryon masses are generated by the scalar fields except for a small explicit mass term equal to $\delta m \sim 150$ MeV for nucleons. The effective
masses decrease at high densities with decreasing scalar fields as the chiral symmetry is partially restored. At low densities
they reproduce the
experimentally known baryon masses:
\begin{eqnarray}
&m^*=g_{i\sigma}\sigma+g_{i\delta}\tau_3\delta+g_{i\zeta}\zeta+\delta m.&
\end{eqnarray}

The coupling constants used to calculate proto-neutron star properties ($g_{N\omega}=11.9$, $g_{N\phi}=0$, $g_{N\rho}=4.03$, $g_{N\sigma}=-9.83$, $g_{N\delta}=-2.34$, $g_{N\zeta}=1.22$,
$k_0=2.37$, $k_1=1.40$, $k_2=-5.55$, $k_3=-2.65$, $k_4=-0.23$, $\epsilon=0.06$, $g_4=38.9$) are chosen to
reproduce the vacuum masses of the baryons and mesons, the nuclear saturation properties
(density $\rho_B=0.15{\rm fm}^{-3}$, binding energy per nucleon $B/A=-16.00$ MeV, compressibility $K=297.32$ MeV),
the asymmetry energy ($E_{sym}=32.50$ MeV), and reasonable values for the hyperon potentials
($U_\Lambda=-29.41$ MeV, $U_\Sigma=20.39$ MeV, $U_\Xi=-10.09$ MeV).
The vacuum expectation values of the scalar mesons are constrained by reproducing the pion and kaon decay constants.

\section{Neutron Stars}

\begin{figure}
\includegraphics[scale=.45]{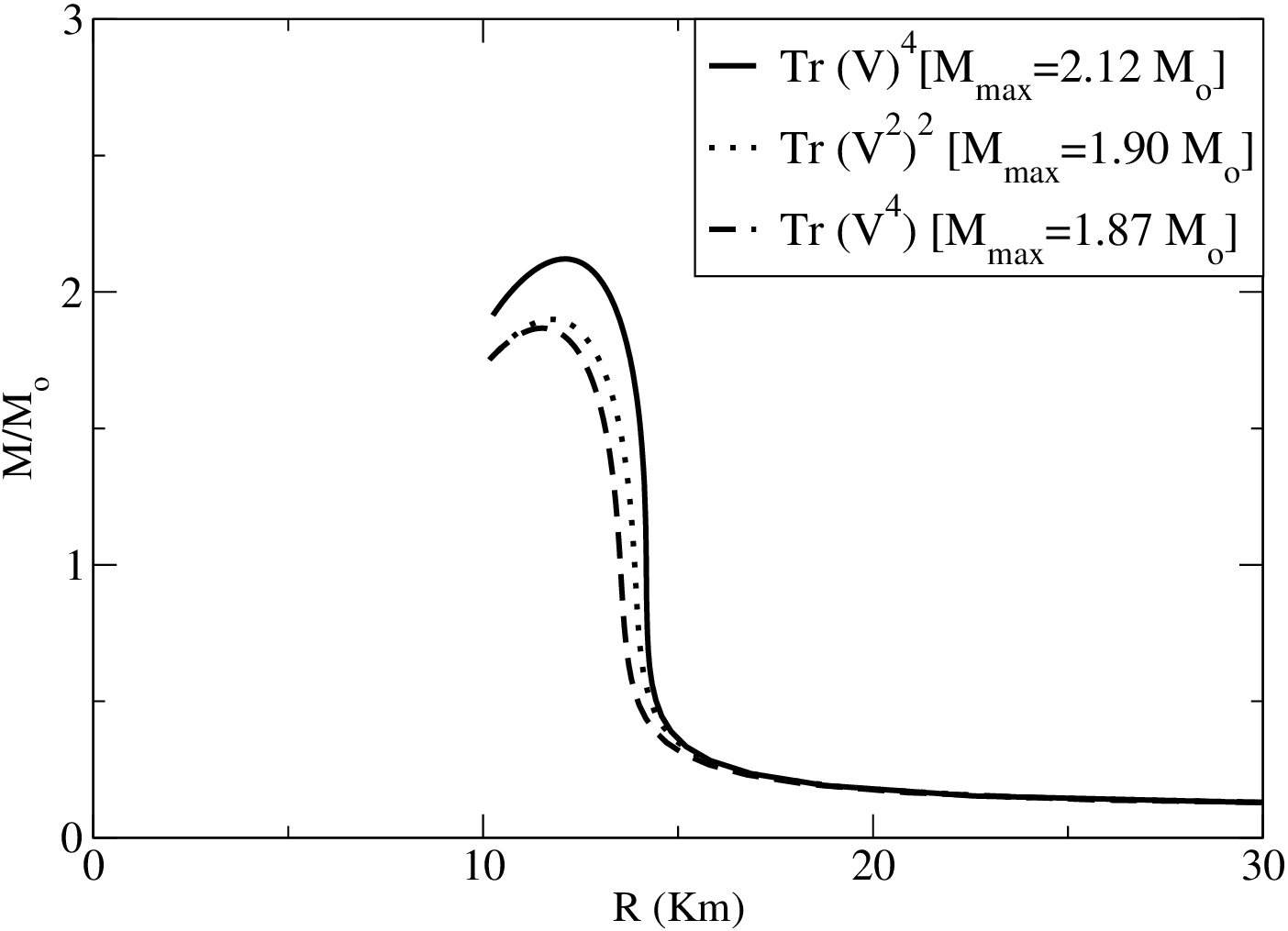}
\caption{Star mass versus radius for different vector meson self interactions.}
\label{massbeta}
\end{figure}

In the calculation of zero-temperature neutron stars the influence of the structure of the
self interaction term of the vector mesons is investigated , as in dense systems baryonic vector densities, and therewith the mean fields of the vector mesons,
become especially important for the equation of state of hadronic matter. The fourth-order self-interaction term $L_{vec,4}$ of the vector mesons
can be written
in different forms in a SU(3)-invariant way. To study the difference in the result three separate coupling schemes are considered,
\begin{equation}
L_{vec,4} =  - g_4 [Tr(V)]^4/4~~~ (a), ~~~ - g_4 [Tr(V^2)]^2~~~ (b),~~~  - g_4 2Tr(V^4) ~~~(c)~~,
\end{equation}
where $V$ stands for the $3$x$3$ matrix of the vector meson
multiplet, which reduces to a diagonal form in the mean field
limit, i. e. $V = ((\omega + \rho)/\sqrt{2},(\omega -
\rho)/\sqrt{2}, \phi)$. The mass-radius relation of the respective
neutron star for the cases (a) to (c) was calculated by solving
the Tolman-Oppenheimer-Volkov (TOV) equations for static spherical
stars (\citet{tov1, tov2}). The results are shown in Fig.
\ref{massbeta} in the SU(2) limit of proton, neutron and electron
matter. A strong coupling of the $\omega$ and $\rho$ meson present in the cases (b) and (c) leads to smaller star masses.
In the following calculations the non-linear coupling (a), which
does not generate a $\rho-\omega$ coupling, is used. This allows
for more massive neutron stars and is also in general agreement
with the observed small mixing of the two mesons.

\begin{figure}
\includegraphics[scale=.45]{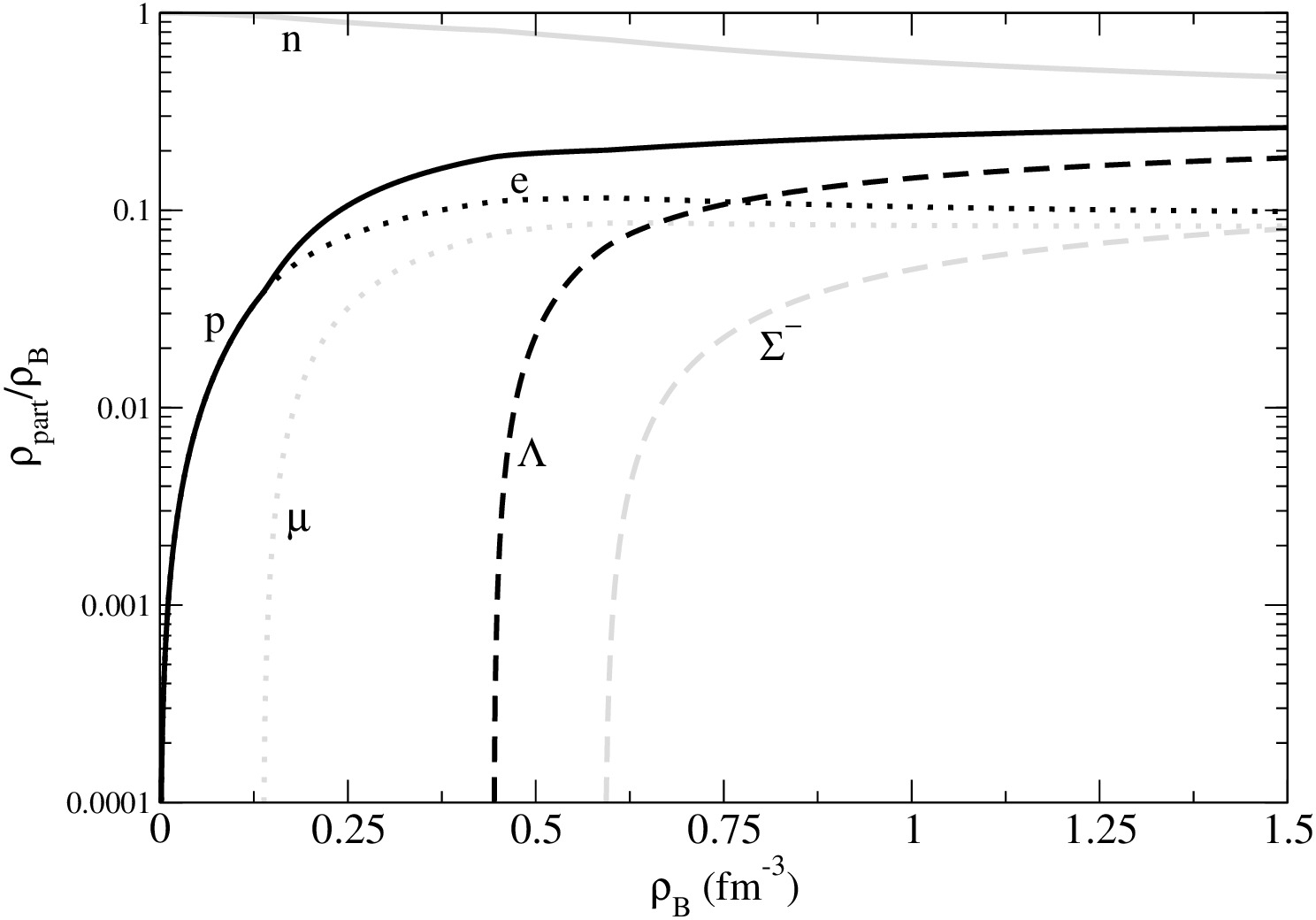}
\caption{The composition of neutron star matter with hyperons.}
\label{pop2}
\end{figure}

\begin{figure}
\includegraphics[scale=.45]{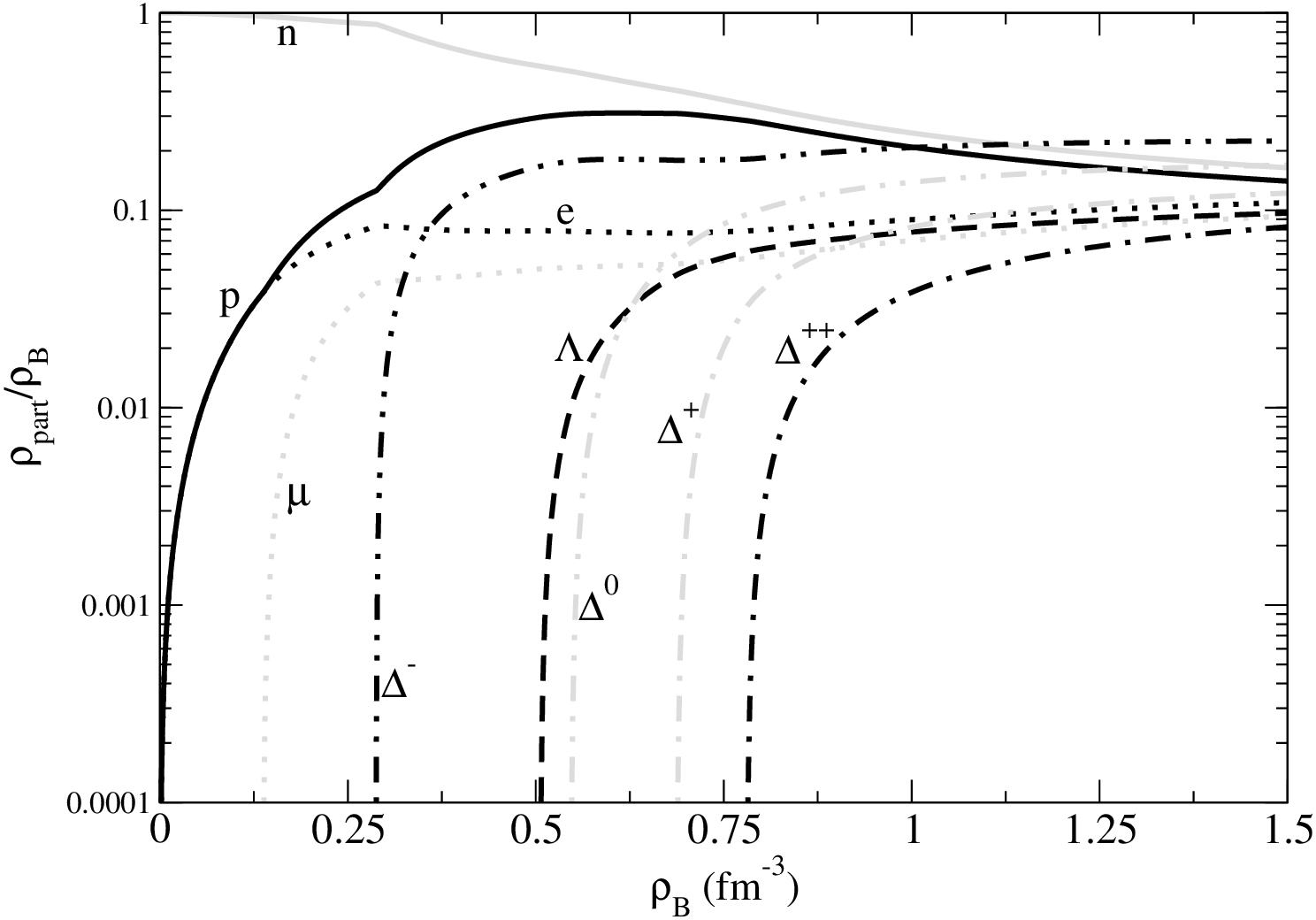}
\caption{The composition of neutron star matter with hyperons and resonances.}
\label{pop3}
\end{figure}

\begin{figure}
\includegraphics[scale=.45]{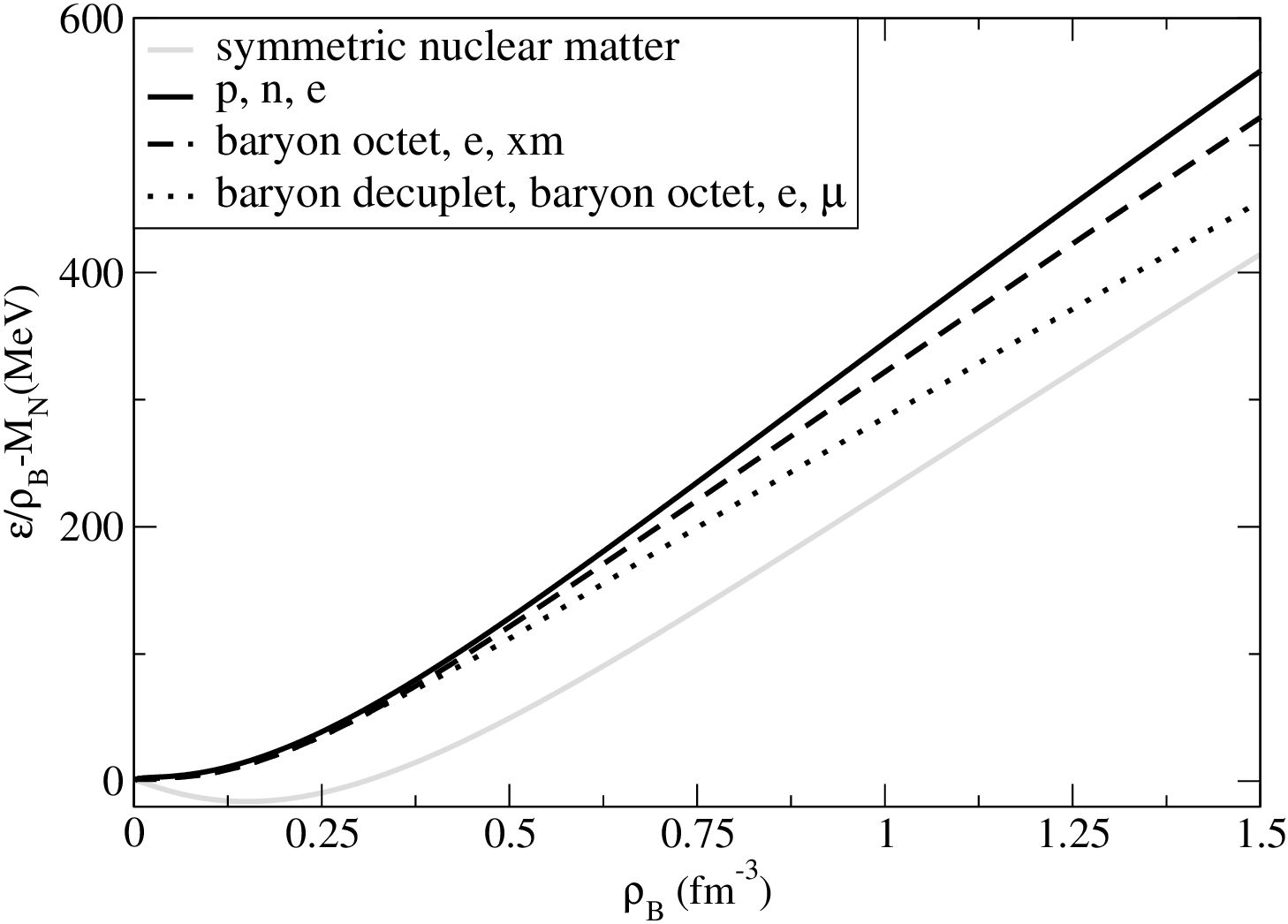}
\caption{Binding energy per nucleon versus baryon density.}
\label{bind}
\end{figure}

The modification of the original parameter set C1 of the model, used in \citet{chiral2} was done in order to investigate the maximum neutron star masses that can be achieved in the model (similar studies in a different approach have been done in \citet{amruta}), while still reproducing
the hadronic masses in the vacuum as well as reproducing the phenomenological values of the basic nuclear matter ground state properties
as listed in the previous section.
Within the model different situations can be analyzed by including
the whole baryon octet or,  in addition, the baryon decuplet. In practice, the only baryons present in the star
besides the nucleons are the $\Lambda$ and $\Sigma^-$ in the first case  (Fig. \ref{pop2}) and
$\Lambda$, $\Delta^{-,0,+,++}$, in the second case (Fig. \ref{pop3}). In the presence of resonances the $\Delta^-$ particle replaces the $\Sigma^-$,
as its effective mass drops faster with density compared to the $\Sigma^-$.
As can be seen from Fig. \ref{bind} the inclusion of new particles, i.e. new degrees of freedom,
softens the equation of state (EOS). The same effect would be observed in the symmetric case, when there is no net isospin instead of no net electric charge,
in the presence of more massive degrees of freedom.
Although the symmetric case has no relevance for neutron stars due to their high intrinsic asymmetry this case is important for example in heavy ion collisions.

\begin{figure}
\includegraphics[scale=.45]{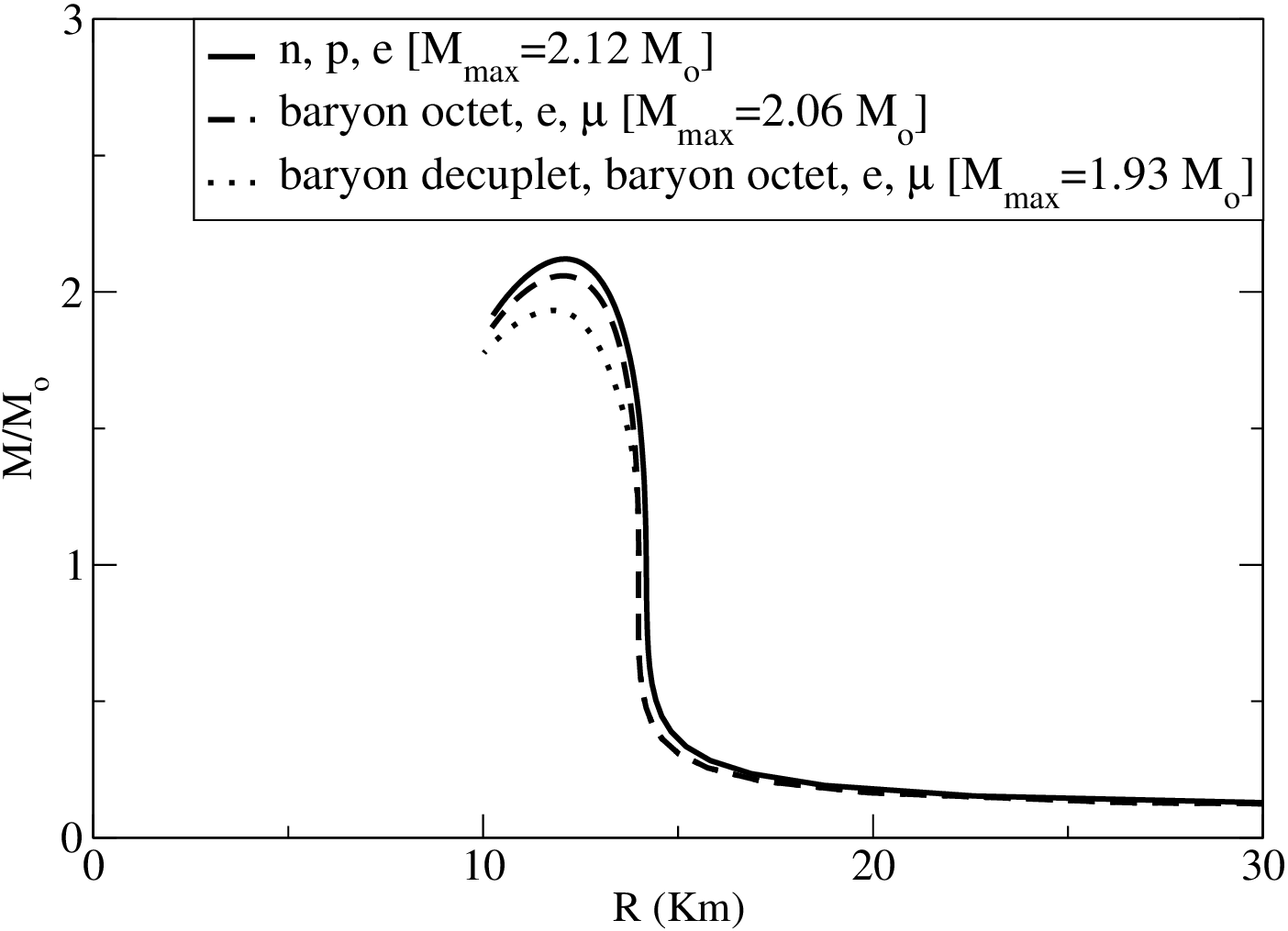}
\caption{Star mass versus radius for different compositions.}
\label{mass}
\end{figure}

The softening of the equation of state causes a decrease of the respective neutron star maximum mass (Fig. \ref{mass}). Besides using the high density EOS for the interior of the star, in order to calculate the mass-radius diagram, an inner crust, an outer
crust and an atmosphere have been considered following \citet{crust}. The studies of the proto-neutron star properties are restricted to the model including the lowest multiplets, i.e. including
the baryon octet, thus avoiding the uncertainties related to the largely unknown coupling strengths of the baryonic decuplet. In this case, it is possible to describe stars with masses higher than $M=2M_{\odot}$ and
still take into account heavy baryonic degrees of freedom.

\begin{figure}
\includegraphics[scale=.45]{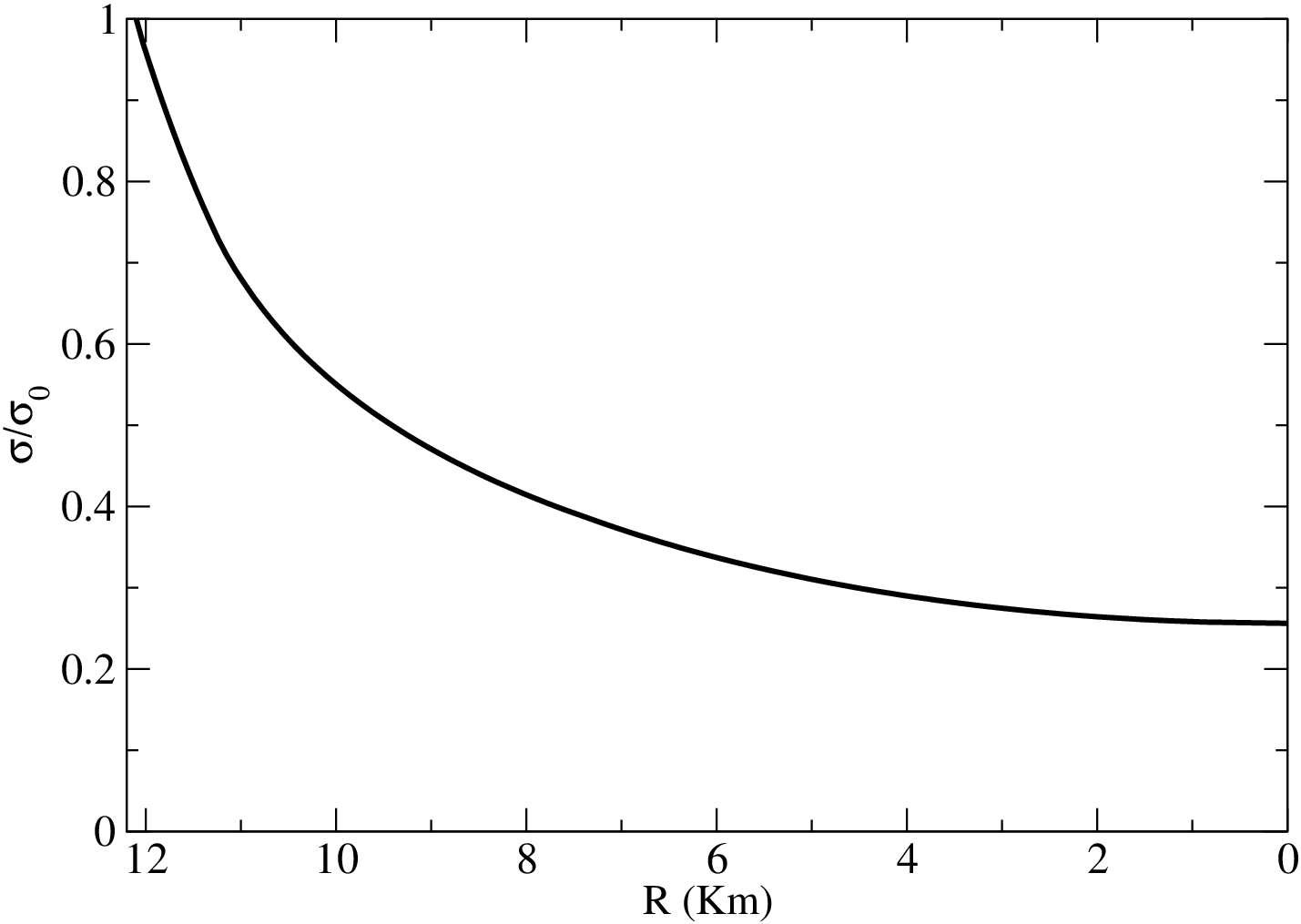}
\caption{Scalar condensate versus star radius.}
\label{raio}
\end{figure}

The transition to the chirally restored phase for any  of the considered self-interactions and sets of baryonic
degrees of freedom
turns out to be a cross over. This happens due to the requirements of beta equilibrium and charge neutrality that make
the different isospin states of baryons with the same vacuum mass appear at different densities of the star, thus smoothing out their effect.
This transition can be seen in the behavior of the scalar condensate $\sigma$ used here as the order parameter for the transition. In Fig. \ref{raio} the condensate was plotted against the star radius showing that in this model the chiral symmetry is partially restored in neutron stars.

\section{Proto-Neutron Stars}

Right after the supernova explosion, due to the neutronization of the matter, the star contains an abundant number of neutrinos that
are trapped in the system. The temperature of the star can reach values of up to 50 MeV.
In order to determine the matter properties under these conditions the thermodynamical potential
of the grand canonical ensemble is solved. It is defined as:
\begin{eqnarray}
&\frac{\Omega}{V}=-L_{scal}-L_{vec}-L_{SB}-L_{vac}\nonumber&\\& \mp T\sum_i \frac{\gamma_i}{(2 \pi)^3}
\int_{0}^{k_{F_i}} \,
d^3k \, \ln(1\pm e^{-\frac{1}{T}(E_i^*(k)-\mu_i^*)}),&
\end{eqnarray}
where $i$ denotes the fermion type (including leptons), $\gamma_i$ the fermionic degeneracy, $E_{i}^* (k) = \sqrt{k^2+{m^*_i}^2}$ the energy, and $\mu_i^*=\mu_i-g_{\omega} \omega_0-g_{\rho} \rho_{03}\tau_3/2-g_{\phi} \phi_0 $
the effective chemical potential (a vanishing chemical potential for muon neutrinos has been assumed here). In the case of leptons $m^* = m$ and $\mu^* = \mu$. The single particle energy is given by
$E_i(k)=E_i^*(k)+g_{\omega} \omega+g_{\rho} \rho \tau_3/2+g_{\phi} \phi$ and the entropy per volume per baryon is defined as $s=S/V/\rho_B=-d\Omega/dT|_{V, \, \mu} / \rho_B$.

\begin{figure}
\includegraphics[scale=.45]{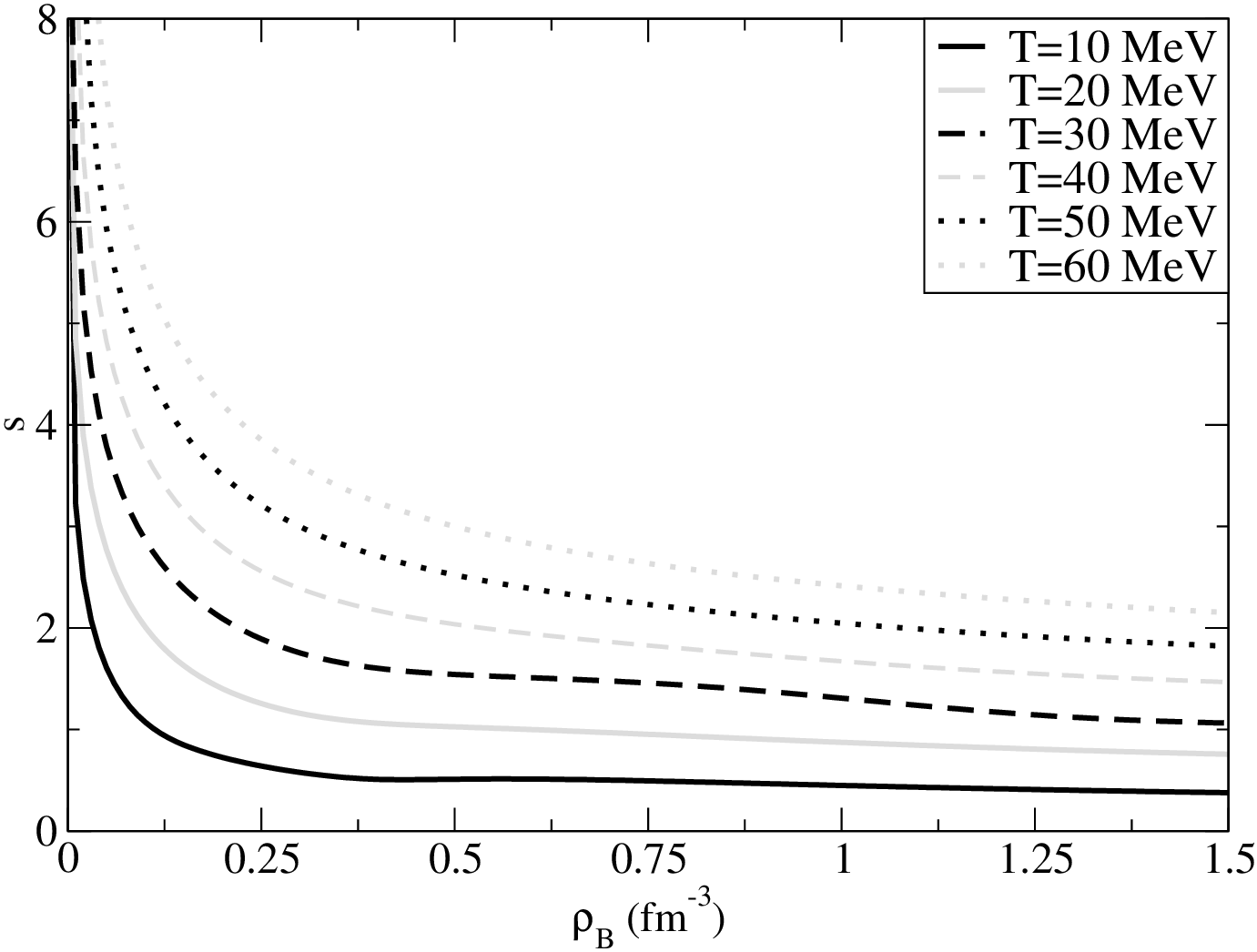}
\caption{Entropy per volume per baryon versus baryon density for different temperatures.}
\label{ent}
\end{figure}

\begin{figure}
\includegraphics[scale=.45]{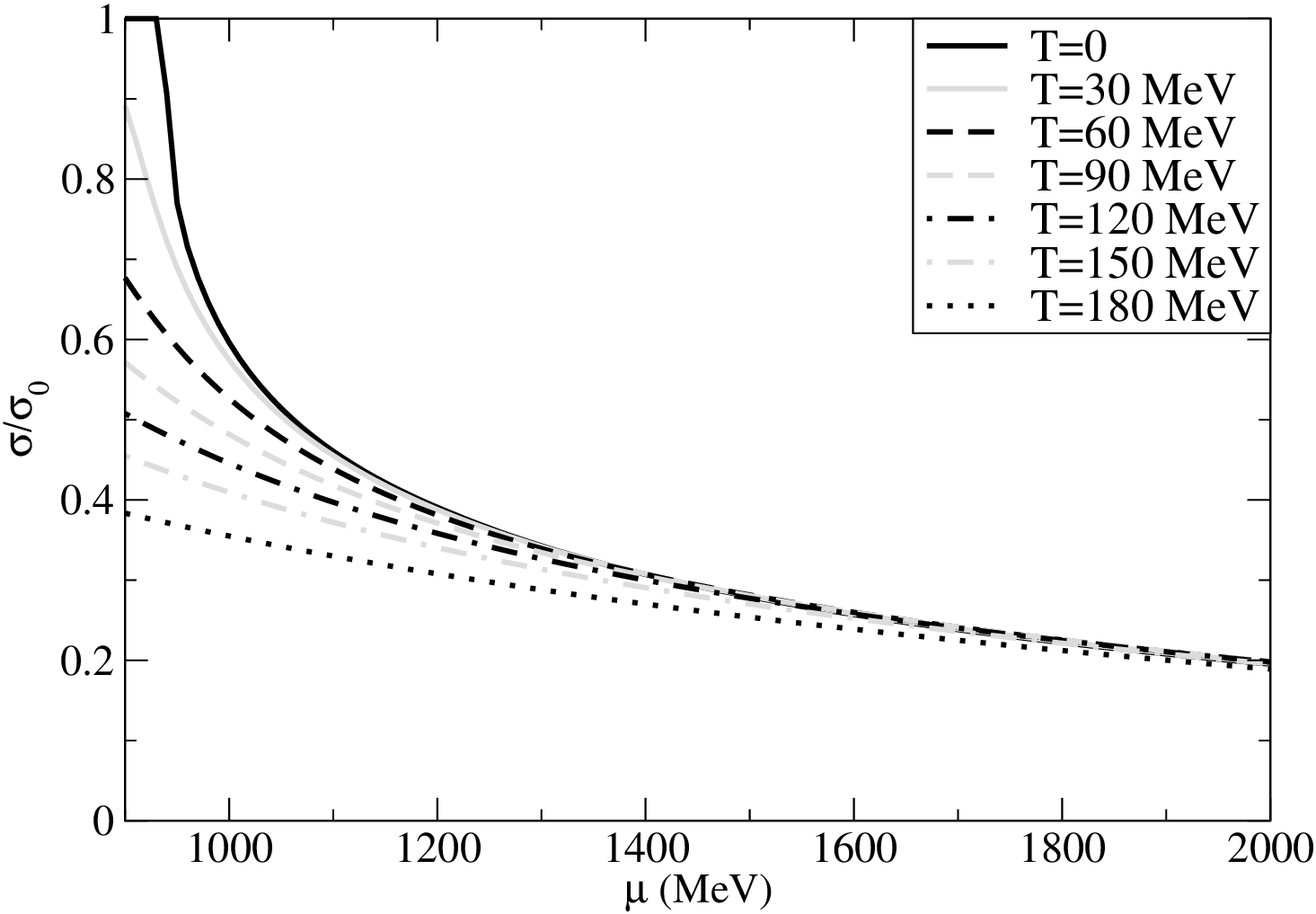}
\caption{Scalar condensate versus chemical potential for different temperatures.}
\label{sigma}
\end{figure}

\begin{figure}
\includegraphics[scale=.45]{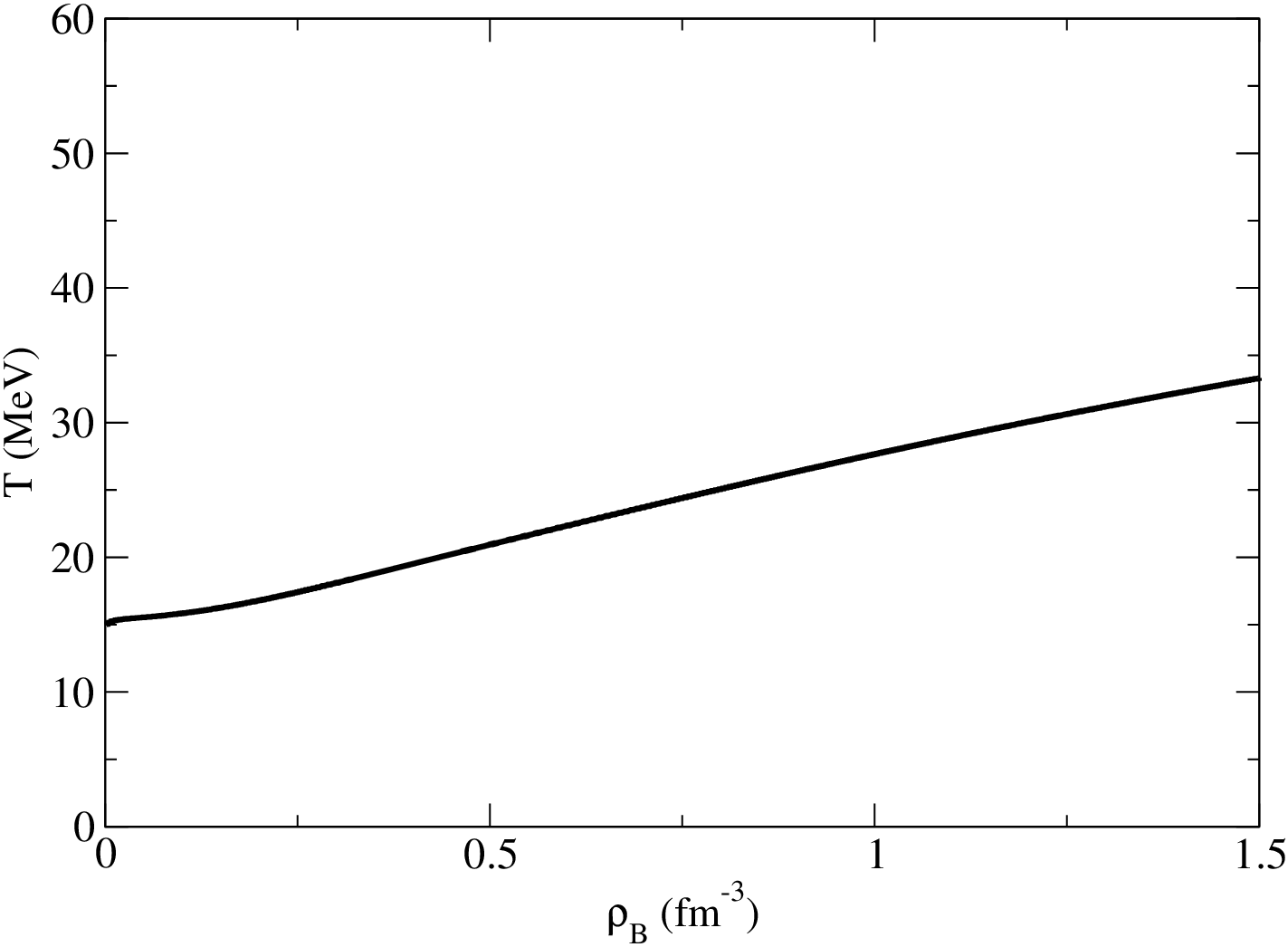}
\caption{Temperature versus baryon density for $T_{\infty}=15$ MeV.}
\label{tinf}
\end{figure}

\begin{figure}
\includegraphics[scale=.45]{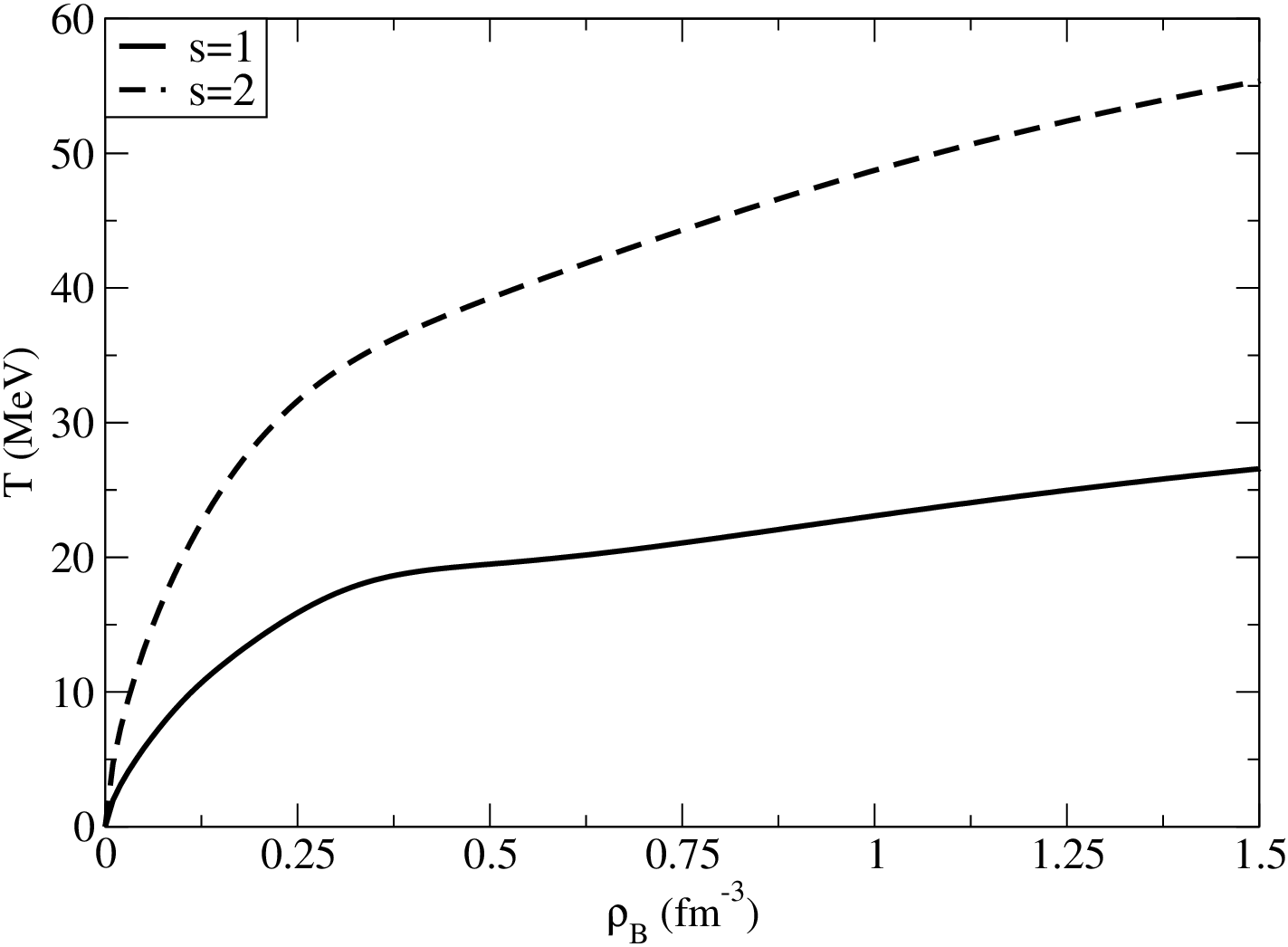}
\caption{Temperature versus baryon density for different entropies per volume per baryon.}
\label{temp}
\end{figure}

To study this complex system two different features are taken into account
separately: finite temperature and high lepton number. In a static approximation of the evolution of the star three different approaches are considered as they have appeared in the literature:

\begin{itemize}
\item constant temperature: in this case the whole star is considered at the same temperature, which is unrealistic,
and the maximum mass of the star is higher for higher temperatures because of the thermal effects on the
binding part of the mass.
In this case the entropy per baryon remains constant with the increase of density except for small densities as can be seen in Fig. \ref{ent}. This effect comes from the fact that even when the baryon density tends to zero, the electron-positron pairs present at finite temperature still contribute for the entropy. The transition to the chirally restored phase becomes smoother
with the increase of temperature since the Fermi surface becomes less important (Fig. \ref{sigma}).

\item metric dependent temperature: in this case, as discussed in \citet{tinf},
the temperature is defined at an infinite distance from the star and the thus defined temperature increases
as the gravitational field created by the presence of the star
mass becomes higher: $T=T_{\infty}/\sqrt{g_{00}}$, where $g_{00}$ is the $00$ component of the
metric tensor.  As the gravitational field increases toward the center of the star the density also
increases as can be seen in Fig. \ref{tinf} for $T_\infty=15$ MeV, but the increase of the temperature from the center to the edge of the
star is not too pronounced. In this case, the maximum mass of the star is higher for higher temperatures $T_\infty$.

\item constant entropy: in this case the star is considered to have a constant entropy throughout, which to some degree agrees with dynamical simulations
of the stellar evolution (\citet{simul}). The temperature is higher in the center of the star and colder at
the edge as can be seen in Fig. \ref{temp} and the maximum mass of the star is higher for higher entropies
also because of the thermal effects on the binding part of the mass.

\end{itemize}

The trapped neutrinos with a chemical potential $\mu_\nu$ are included by fixing the
lepton number defined as $Y_l=(\rho_e+\rho_{\nu_e})/\rho_B$. In consequence
there will be a large number of neutrinos in the star but also an increased
electron density. Therefore, demanding charge neutrality, the proton density increases,
and with higher proton density, the star becomes more isospin symmetric and the neutron Fermi energy decreases.
Thus the increase of lepton number softens the EOS and consequently the maximum mass of the neutron star gets smaller. It is important to keep in mind that this result is dependent on the chosen parameter set and consequently on the particles present in the star. For parameter sets that allow the hyperons to appear at lower densities, the high proton density delays their population allowing the neutron star to be more massive.

These two features can be put together to describe the evolution of the star for the extreme cases: constant temperature and fast increasing temperature throughout the star. After the supernova explosion,
the star is still warm so the temperature is fixed to $T=30$ MeV or the entropy per baryon is fixed to $s=2$
(the temperature increases from $0$ at the edge up to $50$ MeV in the center as in Fig.\ref{temp}).
The star still contains a high abundance of neutrinos that were trapped during the explosion so the lepton
number is fixed to a typical value of $Y_l=0.4$. After several seconds (10-20) the neutrinos escape and $\beta$-equilibrium is established. After about a minute, the temperature of the star has dropped below 1 MeV.

\begin{figure}
\includegraphics[scale=.45]{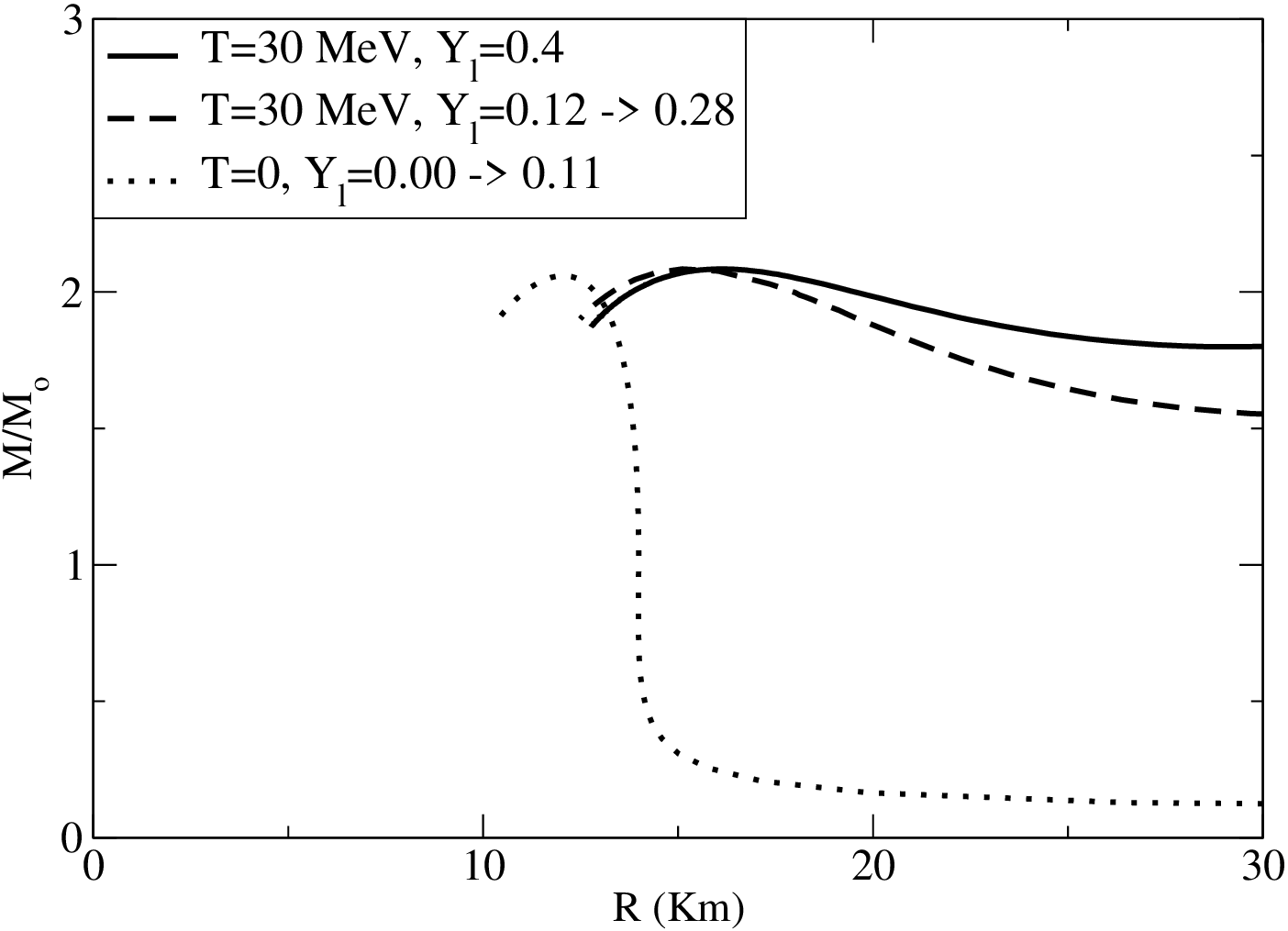}
\caption{Star mass versus radius for different moments of the star evolution defined by temperature and lepton number considering the baryon octet, e, $\mu$.}
\label{MassT}
\end{figure}

\begin{figure}
\includegraphics[scale=.45]{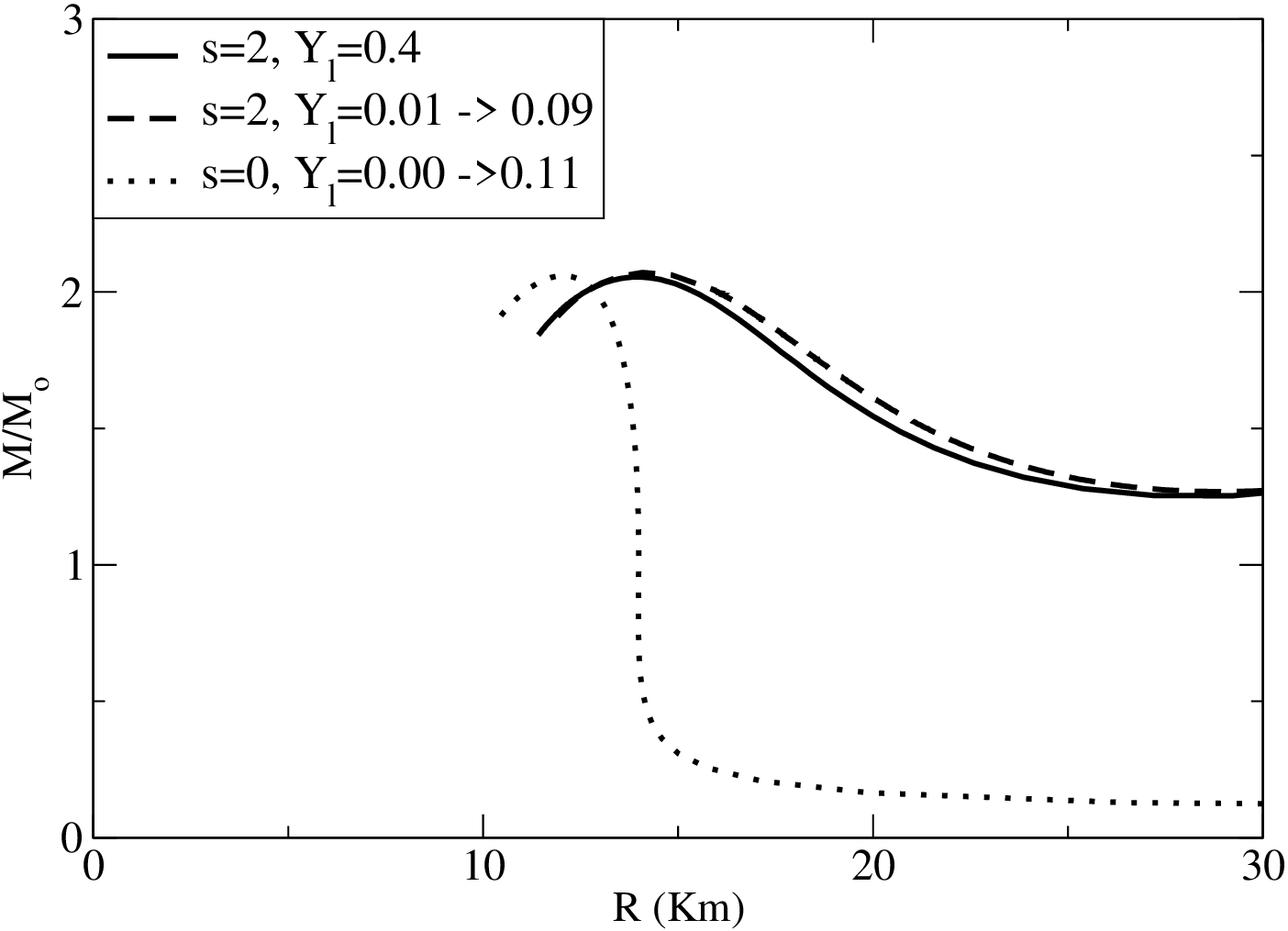}
\caption{Star mass versus radius for different moments of the star evolution defined by entropy and lepton number considering the baryon octet, e, $\mu$.}
\label{finalhyp}
\end{figure}

The balance between the effects of temperature/entropy and
lepton number is very delicate and depends on the parameters of the model.
The first line in table \ref{tabela1} shows the maximum star masses for different entropies and different temperatures.
As can also been seen in Figs. \ref{MassT} and \ref{finalhyp} the intermediate step of the evolution with
$s=2$ and $\mu_\nu = 0$
is more massive than the first one.
But this calculation does not take into account that the baryon number
does not increase as the star gets colder, otherwise it would collapse into a black hole (\citet{baryonmass}).
If the baryon number is fixed starting with the values for the star with $s=2$ or $T=30$ MeV, resp.,  and $Y_l=0.4$, the stable solutions of the colder
cases have a smaller mass than the warmer case as can be seen in the second line of Tab. \ref{tabela1}.

In stars with finite temperature/entropy a crust of high entropy has been used since the shock wave created during the supernova explosion leaves the outer region with a much higher entropy than the rest of the star
and it remains warmer for longer time serving as an insulating
blanket, which delays the star from coming to a complete thermal equilibrium with the interstellar medium (\citet{lattimer}). In this case the crust is stiff enough to generate massive stars for small central energies resulting in big radii. Because of the use of a warm
crust it is also possible to stipulate a minimum mass for each case (fourth line of Tab. \ref{tabela1}). For the assumption of constant temperature the crust used has an entropy $s=5$ so that the inner and outer EOS could be continuously connected
but as a result the minimum mass found is far from a realistic value ($M\sim1.18M_o$, \citet{min}). For the constant entropy case the crust used has a value of $s=4$.

\section{Rotation}

\begin{figure}
\includegraphics[scale=.45]{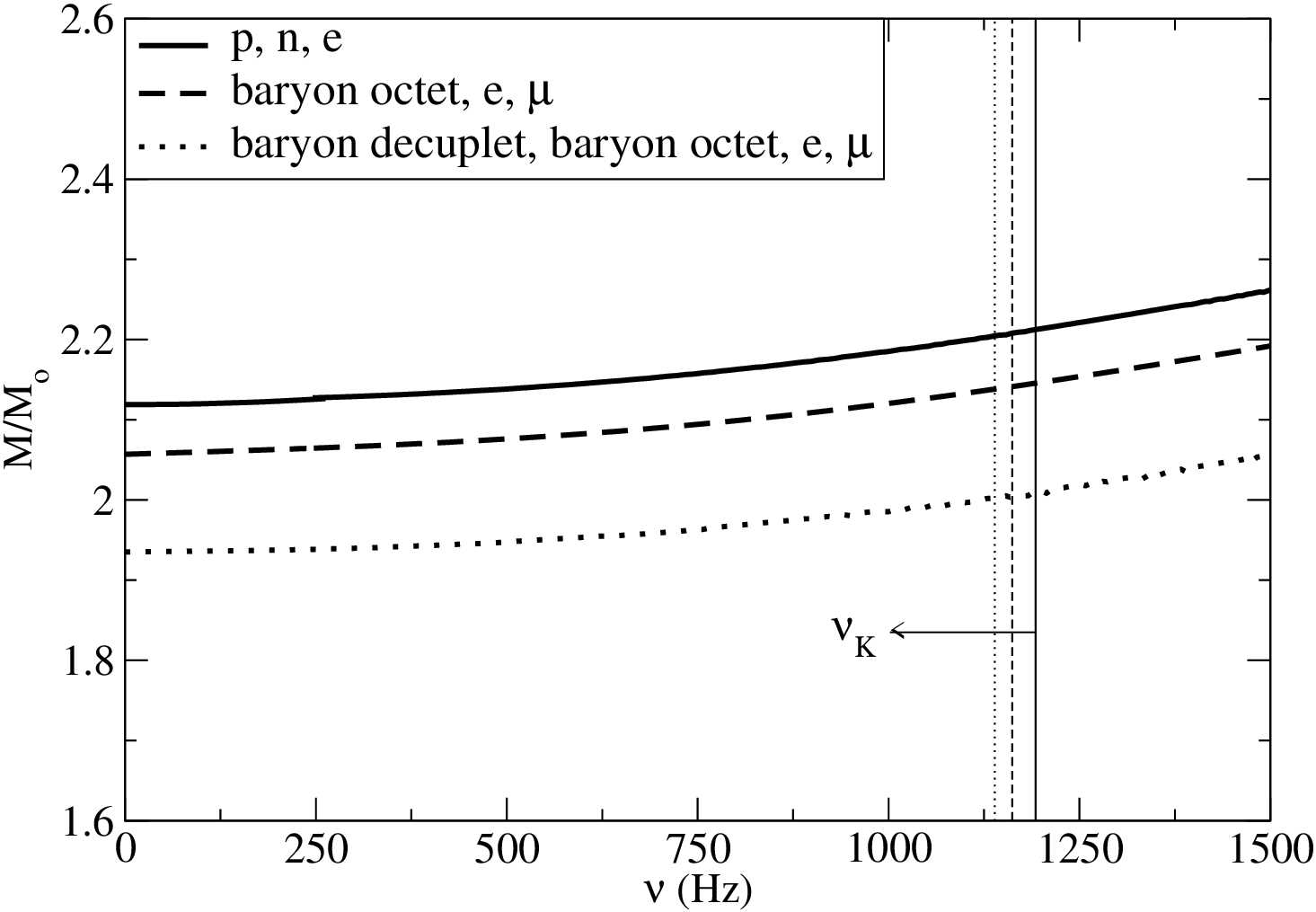}
\caption{Star mass versus rotational frequency of the star for different compositions.}
\label{rot}
\end{figure}

As the proto-neutron star cools down, its possible radii and masses change (Tab. \ref{tabela1}), which means that its moment of inertia changes
and for determining the maximum frequency of the star one has to take into account angular momentum conservation (\citet{baryonmass}).
The maximum frequency (Kepler frequency) with which a star can rotate without starting to
expel matter on the equator, has been determined by including monopole
and quadrupole corrections to the metric due to the rotation and solving the self-consistency equation for $\Omega_K$ as it was
derived in \citet{glendenning}. In the current calculations rotational instabilities, that would potentially reduce the maximum rotational frequency,
are not taken into account. The rotation of the star generates a
modification of the metric and the higher
the rotational frequency, the higher the mass and radius
of the star become (\citet{neutronSchramm2}) as can be seen in Fig. \ref{rot}. The maximum masses calculated in this approximation differ by less than $1\%$ from full calculation results (\citet{rns}) for frequencies around $716$ Hz. For this set of parameters the increase in the maximum masses of
the stars from $\nu=0$ to $\nu=\nu_K$, fixing the baryon number to the value for zero frequency,
is smaller than $5\%$, different from the $15\%$ mass increase in the case that the baryon number is not fixed. But this situation can be identified as the spin down of a cold star with a certain baryon number (in this case $0.28$ x $10^{58}$ N) that continues until it emits all its energy and stops rotating. In this point the star is considered dead.

To consider the rotational effects in the cooling from the very first stage of the neutron star life, besides the baryon number, the angular momentum has to be taken into account
since it remains about constant in the first seconds of the cooling. This new constraint lowers even more the possible maximum mass and Kepler frequency of the star as can be seen in Tab. \ref{tabela2}. As a consequence of that, some stars are stable during a stage of cooling but not in the subsequent one,
causing matter to be lost during the process.

\section{Conclusion}

For the first time an effective model including hyperon degrees of freedom
could generate a neutron star with a mass $M\sim 2.1~M_o$ in agreement with the most massive
one observed but still under discussion (\citet{21}). The inclusion the baryonic spin-3/2 multiplet as additional degrees of freedom still leads
to a quite heavy star of about 1.9 solar masses. This mass might be lowered when the cooling process of the
star is examined and the baryon number is fixed in the early stage of the proto-neutron star
life. The maximum mass of the star increases again when rotation is
included to values beyond $M=2.0 M_o$ for a frequency of $\nu=1122$Hz, the highest frequency measured
for pulsars so far (\citet{fast}) but because of the baryon number and angular momentum constraints, a range of masses in fast rotating stars
leads to unstable states and shedding of matter.

The transition to a largely chirally restored phase takes place in the core of the  neutron
star within the chiral model, although
for all scenarios that were considered here,
the restoration occurs via a
smooth cross over
due to the requirements of beta equilibrium and charge neutrality.

\section{Acknowledgements}

We thank Horst St\"ocker for valuable discussions and suggestions.
V. D. acknowledges the financial support of the
Josef-Buchmann-Stiftung. Part of the numerical work has been done
on the facilities of the Center for Scientific Computing,
University of Frankfurt.

\clearpage

\clearpage

\begin{table}
\begin{center}
\caption{Maximum masses with and without fixing the baryon number, radii
for the maximum masses with fixed baryon number and minimum masses with and
without fixing the baryon number for different moments of the star evolution.\label{tabela1}}
\begin{tabular}{crrrrrr}
\\
\tableline\tableline
stages&$s=2$&$s=2$&$s=0$&$T=30$ MeV&$T=30$ MeV&$T=0$ MeV \\
&$Y_l=0.4$&$\beta$ equil.&$\beta$ equil.&$Y_l=0.4$&$\beta$ equil.& $\beta$ equil.\\
\tableline
$M_{max}(M_o)$&2.05&2.07&2.05&2.08&2.08&2.05\\
fixed $M_B$&2.05&2.01&1.96&2.08&2.03&1.99\\
$R (km)$&14.00&15.87&12.44&16.09&17.23&12.33\\
$M_{min}(M_o)$&1.07&1.07&0.02&2.02&2.02&0.02\\
fixed $M_B$&1.07&1.07&0.94&2.02&2.02&1.77\\
\tableline
\end{tabular}
\end{center}
\end{table}

\begin{table}
\begin{center}
\caption{Maximum frequencies and masses with and without fixing $M_B$ and L for different stages of the cooling.\label{tabela2}}
\begin{tabular}{crrrr}
\\
\tableline\tableline
stages&$s=2$, $Y_l=0.4$&$s=2$, $\beta$ equil.&$s=0$, $\beta equil.$&$s=0$, $\beta equil.$\\
\tableline
$\nu_K(Hz)$&1452.56&1424.48&1515.86&0\\
$M_{max}(M_o)$&2.31&2.33&2.34&2.06\\
\tableline
fixed $M_B$&1228.23&1110.28&1162.03&0\\
&2.25&2.19&2.14&2.06\\
\tableline
fixed L&1195.30&1095.90&1162.03&\\
&2.24&2.19&2.14&\\
\tableline
\end{tabular}
\end{center}
\end{table}


\begin{thebibliography}{}

\bibitem[Kaplan et al.(1986)]{kaplan}
  D.~B.~Kaplan and A.~E.~Nelson,
  %``Strange Goings on in Dense Nucleonic Matter,''
  Phys.\ Lett.\  B {\bf 175}, 57 (1986).
  %%CITATION = PHLTA,B175,57;%%

\bibitem[Pons et al.(2000)]{kcond}
J. A. Pons, S. Reddy, P. J. Ellis, M. Prakash, and J. M. Lattimer,
Phys. Rev. C {\bf 62}, 035803 (2000)


\bibitem[Glendenning et al.(1991)]{glen1}
N. K. Glendenning and S. A. Moszkowski,
Phys. Rev. Lett. {\bf 67}, 2414 (1991)

\bibitem[Glendenning et al.(2001)]{glen2}
N. K. Glendenning,
Phys. Rev. C {\bf 64}, 025801 (2001)

\bibitem[Jha et al.(2006)]{hyperons}
T. K. Jha, P. K. Raina,
P. K. Panda, and S. K. Patra,
Phys. Rev. C {\bf 74}, 055803 (2006)

\bibitem[Jha et al.(2007)]{amruta}
  T.~K.~Jha, H.~Mishra and V.~Sreekanth,
  %``On attributes of a Rotating Neutron star with a Hyperon core,''
  arXiv:0710.5392 [nucl-th].
  %%CITATION = ARXIV:0710.5392;%%



\bibitem[Page et al.(2006)]{review}
D. Page and S. Reddy,
Ann. Rev. Nucl. and Part. Sci., {\bf 56}, 327 (2006).


\bibitem[Papazoglou et al.(1998)]{chiral1}
  P.~Papazoglou, S.~Schramm, J.~Schaffner-Bielich, H.~Stoecker and W.~Greiner,
  %``Chiral Lagrangian for strange hadronic matter,''
  Phys.\ Rev.\  C {\bf 57}, 2576 (1998).

\bibitem[Papazoglou et al.(1999)]{chiral2}
  P.~Papazoglou, D.~Zschiesche, S.~Schramm, J.~Schaffner-Bielich, H.~Stoecker and W.~Greiner,
  %``Nuclei in a chiral SU(3) model,''
  Phys.\ Rev.\  C {\bf 59}, 411 (1999).

\bibitem[Zschiesche et al.(2007)]{phasediagram}
   D.~Zschiesche, G.~Zeeb and S.~Schramm,
  %``Phase structure in a hadronic chiral model,''
  J.\ Phys.\ G {\bf 34}, 1665 (2007).

\bibitem[Hanauske et al.(1999)]{neutronSchramm}
   M.~Hanauske, D.~Zschiesche, S.~Pal, S.~Schramm, H.~Stoecker and W.~Greiner,
  %``Neutron star properties in a chiral SU(3) model,''
  arXiv:astro-ph/9909052.

\bibitem[Schramm et al.(2003)]{neutronSchramm2}
  S.~Schramm and D.~Zschiesche,
  %``Rotating neutron stars in a chiral SU(3) model,''
  J.\ Phys.\ G {\bf 29}, 531 (2003).

\bibitem[Schaffner et al.(1996)]{jurgen}
  J.~Schaffner and I.~N.~Mishustin,
  %``Hyperon rich matter in neutron stars,''
  Phys.\ Rev.\  C {\bf 53}, 1416 (1996).
%  [arXiv:nucl-th/9506011].
  %%CITATION = PHRVA,C53,1416;%%



\bibitem[Tolman(1939)]{tov1}
  R.~C.~Tolman,
  %``Static solutions of Einstein's field equations for spheres of fluid,''
  Phys.\ Rev.\  {\bf 55}, 364 (1939).

\bibitem[Oppenheimer et al.(1939)]{tov2}
  J.~R.~Oppenheimer and G.~M.~Volkoff,
  %``On Massive Neutron Cores,''
  Phys.\ Rev.\  {\bf 55}, 374 (1939).

\bibitem[Baym et al.(1971)]{crust}
  G.~Baym, C.~Pethick and P.~Sutherland,
  %``The Ground State Of Matter At High Densities: Equation Of State And Stellar
  %Models,''
  Astrophys.\ J.\  {\bf 170}, 299 (1971).

\bibitem[Gondek et al.(1997)]{tinf}
  D.~Gondek, P.~Haensel and J.~L.~Zdunik,
  %``Radial pulsations and stability of protoneutron stars,''
  Astron.\ Astrophys.\  {\bf 325}, 217 (1997).

\bibitem[Stein et al.(2006)]{simul}
  J.~Stein and J.~C.~Wheeler,
  %``The Convective Urca Process with Implicit Two-Dimensional Hydrodynamics,''
  Astrophys.\ J.\  {\bf 643}, 1190 (2006).

\bibitem[Takatsuka et al.(1995)]{baryonmass}
  T. Takatsuka, Nucl. Phys. A {\bf 588}, 365 (1995).

\bibitem[Lattimer et al.(1991)]{lattimer}
  J.~M.~Lattimer and F.~D.~westy,
  %``A Generalized equation of state for hot, dense matter,''
  Nucl.\ Phys.\  A {\bf 535}, 331 (1991).

\bibitem[Faulkner et al.(2004)]{min}
  A.~J.~Faulkner {\it et al.},
  %``PSR J1756-2251: a new relativistic double neutron star system,''
  Astrophys.\ J.\  {\bf 618}, L119 (2004).

\bibitem[Glendenning et al.(1994)]{glendenning}
  N.~K.~Glendenning and F.~Weber,
  %``Impact Of Frame Dragging On The Kepler Frequency Of Relativistic Stars,''
  Phys.\ Rev.\  D {\bf 50}, 3836 (1994).

\bibitem[Stergioulas et al. (1995)]{rns}
   N.~Stergioulas and J.~L.~Friedman,
  %``Comparing models of rapidly rotating relativistic stars constructed by two
  %numerical methods,''
  Astrophys.\ J.\  {\bf 444}, 306 (1995)

\bibitem[Nice et al.(2005)]{21}
   D.~J.~Nice, E.~M.~Splaver, I.~H.~Stairs, O.~Loehmer, A.~Jessner, M.~Kramer and J.~M.~Cordes,
  %``A 2.1 Solar Mass Pulsar Measured by Relativistic Orbital Decay,''
  Astrophys.\ J.\  {\bf 634}, 1242 (2005).

\bibitem[Kaaret et al.(2007)]{fast}
  P.~Kaaret {\it et al.},
  %``Discovery of 1122 Hz X-Ray Burst Oscillations from the Neutron-Star X-Ray
  %Transient XTE J1739-285,''
  arXiv:astro-ph/0611716.


\end{thebibliography}
\end{document}